\documentclass[10pt,twocolumn,letterpaper]{article}

\usepackage{cvpr}              %

\usepackage{algorithm}
\usepackage[noend]{algpseudocode}
\usepackage[page]{appendix} %
\usepackage{amsmath}
\usepackage{amssymb}
\usepackage[accsupp]{axessibility}  %
\usepackage{graphicx}
\usepackage{multirow}
\usepackage{booktabs}
\usepackage{siunitx}
\usepackage{cancel}
\usepackage{wrapfig}
\usepackage{xcolor}

\usepackage[pagebackref,breaklinks,colorlinks]{hyperref}

\usepackage[capitalize]{cleveref}
\crefname{section}{Sec.}{Secs.}
\Crefname{section}{Section}{Sections}
\Crefname{table}{Table}{Tables}
\crefname{table}{Tab.}{Tabs.}

\newcommand{\bftab}{\fontseries{b}\selectfont}

\def\cvprPaperID{6517} %
\def\confName{CVPR}
\def\confYear{2023}

\def\R {\mathbb{R}}

\newcommand{\mv}[1]{\mathbf{#1}}

\newcommand{\nvdr}{\textsc{Nvdiffrast}\xspace}

\newcommand{\oldstuff}[1]{}

\newtoggle{preprint}
\toggletrue{preprint}

\begin{document}

\title{Differentiable Shadow Mapping for Efficient Inverse Graphics}

\author{\hspace{-2em}Markus Worchel \hspace{30pt} Marc Alexa\\[10pt]
TU Berlin
}
\maketitle

\begin{abstract}
We show how shadows can be efficiently generated in differentiable rendering of triangle meshes. Our central observation is that pre-filtered shadow mapping, a technique for approximating shadows based on rendering from the perspective of a light, can be combined with existing differentiable rasterizers to yield differentiable visibility information. We demonstrate at several inverse graphics problems that differentiable shadow maps are orders of magnitude faster than differentiable light transport simulation with similar accuracy -- while differentiable rasterization without shadows often fails to converge.
\end{abstract}

\section{Introduction}
\label{sec:intro}

Differentiable renderers have become an essential tool for solving inverse problems in computer vision. They currently come in two flavors: (1) forward rasterization using \emph{local} shading models~\cite{LoperBlack:2014:OpenDR, Chen:2019:dibr, Chen:2021:dibrpp} and (2) path tracing and/or Monte Carlo methods for \emph{global} light transport simulation~\cite{Li:2018:Redner, Nimier:2019:Mitsuba2, Zhang:2020:PathSpaceDiff, Jakob:2022:Mitsuba3}. While local methods are orders of magnitude faster, they lack effects of global light interaction such as shadows, caustics, or indirect illumination. 

\begin{figure}
    \centering
    \includegraphics[width=\linewidth]{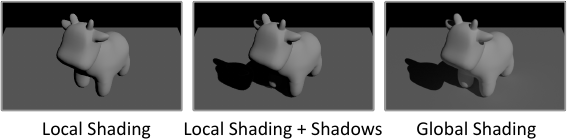}
    \caption{Adding shadows (middle) to local shading (left) is a significant step towards global light transport simulation (right).}
    \label{fig:local_global_shading}
\end{figure}

Modern methods in real-time graphics can generate surprisingly realistic images by using efficient approximations of global effects. The single most important aspect for increasing the realism of local shading is the consideration of \emph{shadows} (see Figure~\ref{fig:local_global_shading}): 
for each pixel to be shaded, check if the path to a light source is unobstructed before evaluating a local shading model. 
Doing this accurately is costly and many approximate techniques have been developed. Our central observation is that one of the oldest and most widely used, \emph{shadow maps}~\cite{Williams:1978:ShadowMapping} (see Section~\ref{sec:sm}), can be adapted to work in differentiable rendering frameworks.

\begin{figure}
    \centering
    \includegraphics[width=\linewidth]{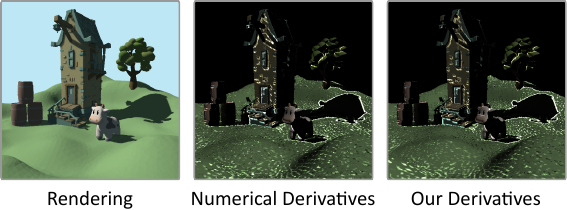}
    \caption{Complex scene (330k triangles) rendered in real time with our differentiable shadow mapping. It shows self-shadowing of objects, shadowing between objects, and colored surfaces. Finite differences and our automatic derivatives w.r.t.\ movement of the light (note the non-zero derivatives at shadow boundaries).
    }
    \label{fig:complex_scene}
\end{figure}

In Section~\ref{sec:dif} we explain how (certain approximations of) shadow mapping can be differentiated, exploiting existing differentiable rasterizers. Our main idea is similar to shadow maps: exploit efficient rasterization from the light's point of view. For differentiable shadows this means:
Existing differentiable rasterizers handle discontinuities of primary visibility along primitive borders; 
\begin{wrapfigure}{r}[12pt]{0.25\linewidth}
  \hspace{-15pt}
  \includegraphics[width=\linewidth]{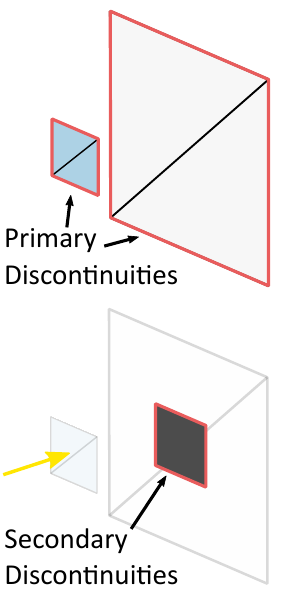}
  \vspace{-10pt}
\end{wrapfigure}
we use this machinery to handle discontinuities of secondary visibility along shadow borders.
The resulting images contain shadows and are differentiable (see Figure~\ref{fig:complex_scene}). For many inverse graphics problems the level of realism they provide will suffice, while being generated significantly faster than with global methods. This is important for machine learning tasks, where the renderings (and their derivatives w.r.t.\ scene parameters) are computed repeatedly. We provide details of the implementation and how parameters affect optimization based on target images in Section~\ref{sec:implementation}.

Given the importance of shadows for realistic image synthesis, it is unsurprising that many inverse problems heavily depend on them. We demonstrate the trade-off and possibillities of differentiable shadow mappings in several applications, ranging from pose estimation, over different types of geometry reconstruction, to interactive editing by manipulating shadows (Section~\ref{sec:applications}).

The main idea of this work, using existing differentiable rendering frameworks that readily resolve the problem of visibility discontinuities for dealing with discontinuities of secondary rays, may be useful for various other scenarios beyond shadows. We elaborate on this and other immediate consequences of our approach in Section~\ref{sec:conclusions}.

\section{Related Work: Differentiable Rendering}
\label{sec:related}

Differentiable rendering enables computing derivatives of the rendering process, which generates a virtual image from a set of scene parameters. While there are differentiable renderers for a variety of scene representations, including volumes~\cite{NimierDavid:2022:InverseVolumeRendering} or implicit surfaces~\cite{Bangaru:2022:NeuralSDFReparam, Vicini:2022:DiffSdf, Jiang_2020_CVPR}, we focus on triangle meshes here and refer to Kato~\etal{}~\cite{Kato:2020:DiffRenderingSurvey} for a broader overview. 

Many differentiable mesh renderers are based on rasterization, effectively point-sampling the scene on a regular grid in the image plane. Point sampling means that a primitive either covers a pixel or not; this binary outcome makes computing derivatives w.r.t.\ scene parameters difficult. Some differentiable rasterizers approximate the gradient computation~\cite{LoperBlack:2014:OpenDR, Kato:2018:NMR} whereas others approximate the rasterization~\cite{Liu:2019:SoftRas, Chen:2019:dibr, Chen:2021:dibrpp, Petersen:2022:gendr, Takimoto:2022:Dressi, deLaGorce:2008:DEODR, Laine:2020:nvdiffrast}, using a form of smoothing or anti-aliasing to ensure differentiability. Approaches also differ in terms of \emph{local} shading of a pixel, from considering only given colors to evaluating reflection models based on the BRDF.

At the other end of the spectrum are complex physically-based differentiable renderers~\cite{Li:2018:Redner, Nimier:2019:Mitsuba2, Zhang:2020:PathSpaceDiff, Jakob:2022:Mitsuba3}, often based on Monte Carlo integration to numerically approximate the rendering equation~\cite{Kajiya:1986:RenderingEq}. For handling visibility discontinuities in the gradients, several techniques have been developed~\cite{Li:2018:Redner, Loubet:2019:Reparametrizing, Bangaru:2020:was, Zhou:2021:VectorizationDiffVis}.
Significant effort has been made to improve the immense time and memory requirements~\cite{NimierDavid:2020:RB, Vicini:2021:PRB, Jakob:2022:DrJit}. While modern GPUs possess specialized hardware for ray-triangle intersection, they still cannot match the performance of rasterization-based renderers.

Multi-view 3D reconstruction and related inverse problems such as view synthesis and material estimation are prime examples for the potential of differentiable rendering. 
Recent combinations of inverse graphics with machine learning techniques, like NeRF~\cite{mildenhall2020nerf, Barron:2021:MipNerf, Barron:2022:MipNerf360, Niemeyer:2021:Regnerf, Zhang:2021:NeRFactor} or neural implicit surfaces~\cite{Niemeyer:2020:DVR, Oechsle:2021:UNISURF, yariv:2021:volsdf, wang:2021:neus, Zhang:2021:physg, Yariv:2020:idr} are particularly successful.
It is still common to use local shading models, especially in combination with efficient differentiable rasterizers for triangle meshes. Missing shadows or indirect illumination in the differentiable rendering process are often compensated for by assuming a setting where these effects are either neglegible (\eg{} co-located camera and light~\cite{Luan:2021:Unified, Zhang:2022:IRON} 
or soft environment lighting~\cite{Munkberg:2022:nvdiffrec, Hasselgren:2022:nvdiffrecmc}),
by modeling the interaction of light and material as a black box neural network~\cite{Worchel:2022:NDS}, or simply ignored~\cite{Hasselgren:2021:nvdiffmodeling, Rozumnyi:2021:ShapefromBlur}. In addition to analysis-by-synthesis settings, differentiable renderers have been used in a variety of contexts to train machine learning models~\cite{Pavllo:2020:ConvGenerationTexturedMeshes, Chen:2019:dibr, Huang:2020:ARCH, Liu:2021:NeuralActor, Wu:2020:ProbablySymmetric, Michel2022:Text2Mesh, Rueckert:2022:ADOP, wang:2020:self6d, zhang:2020:imagegans, Habermann:2021:DeepDynamicCharacters, Lattas:2022:AvatarMe}.

Despite the popularity in computer graphics, surprisingly little effort has been spent on combining local shading models and approximate global solutions in the context of differentiable rendering. 
Lyu \etal{}~\cite{Lyu:2021:EfficientDiffShadow} take a first step in this direction by approximating soft shadows produced by image-based lighting, however, their gradients are so far limited by the proxy geometry~\cite{Ren:2006:SoftShadows} used for shadow computation. 
In contrast, we use the full mesh geometry for the shadows and are not restricted to a specific class of deformations.

\section{Background: Shadow Mapping}
\label{sec:sm}

Shadow mapping~\cite{Williams:1978:ShadowMapping} is based on the observation that points in a scene are lit by a (point) light source if they are closest to the light source. The \emph{shadow map} is a depth image from the perspective of the light source. If a point in the scene is further from the light than the value stored in the shadow map, it is in the shadow. Otherwise it has the value stored in the shadow map, is closest to the light source, and lit. In local shading models, light is added over all light sources and the shadow computation is performed for each individual light, using a shadow map per light. In the following we consider a single light source and shadow map. 

Formally, let the scene be given as the set $\mathcal{X} \subset \R^3$. A point light source is defined by its position $\mv{p}$. We assume a projective model and represent positions in homogeneous coordinates, so $\mv{p}$ can be at infinity, creating a directional light source, in which case $\mv{p}$ is the direction vector. Denote the distance of $\mv{x} \in \R^3$ to the light source as $d: \R^3 \mapsto [0:d_{\max}]$,
where the distance to a directional light source is measured from a plane outside of $\mathcal{X}$ so that distance values are finite and non-negative.

\begin{wrapfigure}{r}[12pt]{0.2\linewidth}
\vspace{-10pt}
\hspace{-15pt}
  \includegraphics[width=\linewidth]{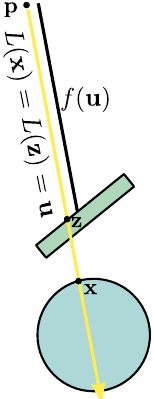}
  \label{fig:primary_seconday}
  \vspace{-10pt}
\end{wrapfigure}
The shadow map is parameterized over the space of rays emanating from $\mv{p}$. For simplicity we assume that the rays from $\mv{p}$ to any point in the scene can be parameterized over a single plane and we assume a rectangle $\mathcal{U}$ on this plane to be the parameter space for the rays; if the light source is placed so that this condition is violated, several planes (\ie{}, arranged in the form of a box) may be used. The mapping from scene points $\mv{x} \in \R^3$ to $\mathcal{U}$ is the linear projective camera transformation $L: \R^3 \mapsto \R^2$. In this setup, the shadow map can be expressed as
\begin{equation}
    f: \mathcal{U} \mapsto \R,
    \quad
    f(\mv{u}) = \min
    \left\{
    {d(\mv{x}): L(\mv{x}) = \mv{u}}, \mv{x} \in \mathcal{X}
    \right\}
\end{equation}
Given the shadow map, we get the \emph{visibility} of any point in the scene w.r.t.\ the light source as
\begin{equation}
    V(\mv{x}) =
    \begin{cases}
        0 & d(\mv{x}) > f(L(\mv{x}))\\
        1 & \mathrm{else}.%
    \end{cases}
\end{equation}
It is important to note that the `else' case really reduces to $d(\mv{x}) = f(L(\mv{x}))$ because $f$ had been constructed to store the smallest possible value of $f$ in direction $L(\mv{x})$. 

\begin{figure}
    \centering
    \includegraphics[width=\linewidth]{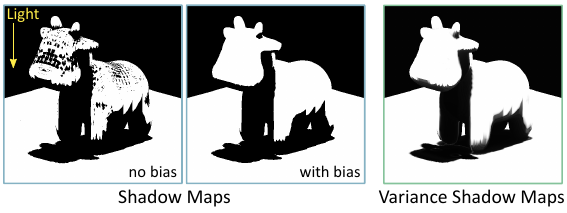}
    \vspace{-0.65cm}
     \caption{
     Shadow mapping suffers from self-shadowing artifacts (``shadow acne") that are alleviated by biasing the depths towards the light source in the visibility test. Variance shadow mapping handles these artifacts implicitly.
     }
    \label{fig:shadow_mapping_comparison}
\end{figure}

The usefulness of shadow maps arises from the fact that they can be efficiently approximated using graphics hardware: simply render the scene from the perspective of the light source, disabling all lighting computations and storing only the depth values. However, a naive implementation will suffer from severe artifacts. First, the condition for a point to be lit is $d(\mv{x}) \le f(L(\mv{x}))$ in finite precision arithmetic, which may or may not happen for visible points on a smooth surface because of the different discretization of the rendered image and the shadow map. The resulting problem is referred to as ``shadow acne" and commonly tackled by introducing a \emph{bias} $\beta$ for the depth comparison, considering points with $d(\mv{x}) \le f(L(\mv{x})) + \beta$ to be lit (Figure~\ref{fig:shadow_mapping_comparison}). While suitably chosen bias attenuates artifacts along smooth surfaces, it cannot remedy aliasing across shadow boundaries. 

Various techniques have been suggested for dealing with aliasing artifacts resulting from the discrete nature of shadow maps~\cite{Annen:CSM,Annen:ESM,Donnelly:2006:VSM,Peters:MSM}. Many such approaches can be traced back to or interpreted as approximations of \emph{percentage closer filtering}\cite{Reeves:1987:percentagecloser}. Rather than comparing $d(\mv{x})$ to only $f(L(\mv{x}))$ we consider a small neighborhood around $L(\mv{x})$ in the shadow map and take the percentage of the rays around $L(\mv{x})$ that would lit $\mv{x}$. In a slight generalization of the original formulation, we assume the neighborhood to be implicitly defined by a smooth (compactly supported) kernel $k_{\mv{x}}(\mv{u})$ satisfying $\int k_{\mv{x}}(\mv{u})\, \mathrm{d}\mv{u} = 1$. Then the desired visibility function is
\begin{equation}
\label{eq:pcf}
    v(\mv{x}) =
    \int_{d(\mv{x}) \le f(\mv{u})} k_{\mv{x}}(\mv{u})\, \mathrm{d}\mv{u}. 
\end{equation}
Note that in this case we can indeed have $d(\mv{x}) < f(\mv{u})$ because $\mv{u}$ is not restricted to $L(\mv{x})$ but varies over a region. The visibility function is taking on values in $[0,1]$, and in particular the boundaries of shadows are smoothly transitioning between the two extreme values. This also means that on completely lit smooth surfaces, even if the visibility is not 1 in concave areas, it is always smoothly varying and close to 1, eliminating the necessity to estimate a suitable bias, which can be difficult for some scenes~\cite{10.1145/2556700.2556706}.

Percentage closer filtering successfully alleviates many sampling artifacts but is costly: for every point $\mv{x}$ in the scene the integral needs to be evaluated. One would much rather \emph{pre-filter} the shadow map~\cite{Annen:CSM}. Yet, simply convolving $f$ with $k$ and then evaluating the smoothed shadow map is different. This can be readily seen as the result would take only the discrete values 0 and 1. Better approximations (see \cite{hasenfratz:inria-00281388} for an overview) generally pre-warp $f$ and/or pre-filter not only $f$ but also simple functions $g(f)$, such as $g(f) = f^k$ or $g = \exp$. Then an approximation to Eq.~\eqref{eq:pcf} is built from the set of pre-computed functions. 

While our approach works with all such approximations, we demonstrate it at the example of \emph{variance shadow maps}~\cite{Donnelly:2006:VSM}. The approximation is built by taking the view that the local shadow map may be considered a probability distribution. This makes sense because it is irrelevant what points are in the integration domain in Eq.~\eqref{eq:pcf} as long as the measure of the integration domain is correct. The local weighted mean and variance of the shadow map are
\begin{equation}
\label{eq:meanvar}
\mu_{\mv{x}} = \int k_{\mv{x}}(\mv{u})f(\mv{u})\, \mathrm{d}\mv{u},
\quad
\sigma^2_{\mv{x}} = \int k_{\mv{x}}(\mv{u})f(\mv{u})^2\, \mathrm{d}\mv{u} - \mu_{\mv{x}}^2.
\end{equation}
We note that $v(\mv{x})$ is a measure and since the shadow map $f$ is positive the best bound for this measure is given by the \emph{one-sided Chebyshev} or \emph{Cantelli} inequality, yielding
\begin{equation}
    v(\mv{x}) \le 
    \frac{\sigma_{\mv{x}}^2}
    {\sigma_{\mv{x}}^2 + (d(\mv{x})-\mu_{\mv{x}})^2}.
\end{equation}
Notice that, similar to percentage closer filtering, it may happen that $d(\mv{x}) < \mu_{\mv{x}}$ because $\mu_{\mv{x}}$ is a filtered version of $f$. Since in this case the approximation of the visibility function would decrease as $d(\mv{x})$ comes closer to the light source, Donnelly and Lauritzen~\cite{Donnelly:2006:VSM} suggest the following slightly modified visibility function:
\begin{equation}
    \hat{v}(\mv{x}) = 
    \begin{cases}
    \frac{\sigma^2_{\mv{x}}}
    {\sigma^2_{\mv{x}} + (d(\mv{x})-\mu_{\mv{x}})^2},
    &
    d(\mv{x}) > \mu_{\mv{x}}\\
    1 & \mathrm{else}.
    \end{cases}
\end{equation}
The resulting shadow maps (see Figure~\ref{fig:shadow_mapping_comparison}) have been observed to be very similar to percentage closer filtering.
\section{Differentiable Filtered Shadow Mapping}
\label{sec:dif}

Our goal is to differentiate the visibility function with respect to the scene parameters $\Theta$. The binary visibility $V(\mv{x})$ has zero derivative almost everywhere. The exact form $v(\mv{x})$ of percentage closer filtering (Eq.~\eqref{eq:pcf}) is smooth, yet the integration domain not only varies with the scene parameters but also depends on $\mv{x}$. We suggest to rather consider $\hat{v}$.

The main difficulty in computing $\frac{\partial \hat{v}}{\partial \Theta}$ is that the shadow map $f$ is only piecewise smooth, and the boundaries of the smooth regions may change with $\Theta$. This makes the derivatives of the integrals in Eq.~\eqref{eq:meanvar} challenging to compute. Similar in spirit to the original shadow map idea~\cite{Williams:1978:ShadowMapping}, we can use existing rasterizers. While the original motivation was efficiency, and this is also true in our case, an additional point is that significant effort has been spent to estimate such derivatives when rendering from the perspective of the viewer~\cite{Laine:2020:nvdiffrast, Liu:2019:SoftRas, LoperBlack:2014:OpenDR, Kato:2018:NMR, Chen:2019:dibr, Chen:2021:dibrpp, deLaGorce:2008:DEODR}. We can readily exploit any existing solution and simply render from the perspective of the light. Most differentiable rasterizers effectively replace the piecewise smooth functions $f$ and $f^2$ with smoothed functions $\tilde{f}$ and $\tilde{f^2}$.
Attenuating aliasing artifacts in pre-filtered shadow mapping \emph{requires} smoothing, so the smoothing by differentiable rasterizers is complementary in our setting. %

In the following, the vectors $\tilde{\mv{f}}$ and $\tilde{\mv{f}^2}$ represent the discrete versions of $\tilde{f}$ and $\tilde{f^2}$, with $\nabla \tilde{\mv{f}}$ and $\nabla \tilde{\mv{f}^2}$ their Jacobians w.r.t.\ $\Theta$. The elements in $\hat{v}$ dependent on the shadow map are $\mu$ and $\sigma^2$. The weighted mean $\mu$ is a convolution that, in the discrete case, can be expressed as the scalar product
\begin{equation}
    \mv{m}_1 = \mv{G} \tilde{\mv{f}}, \label{eq:moment1}
\end{equation}
where $\mv{G}$ is a Toeplitz matrix and $\mv{m}_1$ is the first discrete moment of the shadow map. The derivative is then simply
\begin{equation}
    \frac{\partial \mv{m}_1}{\partial \Theta}
    = 
    \frac{\partial \mv{m}_1}{\partial \tilde{\mv{f}}}
    \frac{\partial \tilde{\mv{f}}}{\partial \Theta}
    = 
    \mv{G} \nabla \tilde{\mv{f}}.
\end{equation}
For $\sigma^2$ we introduce the second discrete moment
\begin{equation}
    \mv{m}_2 = \mv{G} \tilde{\mv{f}^2}, \label{eq:moment2}
\end{equation}
which similarly has the derivative 
\begin{equation}
    \frac{\partial \mv{m}_2}{\partial \Theta}
    = 
    \frac{\partial \mv{m}_2}{\partial \tilde{\mv{f}^2}}
    \frac{\partial \tilde{\mv{f}^2}}{\partial \Theta}
    =
    \mv{G} \nabla \tilde{\mv{f}^2}.
\end{equation}
The discrete image of $\sigma^2$ can be computed from the %
moments as $\mv{m}_2-\mv{m}_1 \otimes \mv{m}_1$ and its derivative directly follows from expressions above and the product rule for $\mv{m}_1 \otimes \mv{m}_1$. 

The distance $d(\mv{x})$ is independent of the shadow map. Its variation with $\Theta$ stems from possible modification of the position of the light source or geometry and can be easily expressed. Lastly, $\hat{v}$ is a piecewise rational function of these variables, with the boundary between the two pieces at $d = \mu$ depending on both $d$ and $\mu$. This requires no special treatment, however, because $\hat{v}$ is a continuously differentiable function across $d=\mu$, as we see from
\begin{equation}
    \frac{\partial \hat{v}}{\partial{\mu}} = 
    -\frac{\partial \hat{v}}{\partial{d}} = 
    \begin{cases}
    2 \frac{\sigma^2 ( d-\mu)}
    {\left(\sigma^2 + ( d-\mu)^2\right)^2}
    &
    d > \mu\\
    0 & \mathrm{else}
    \end{cases}.
\end{equation}
Thus, $\frac{\partial \hat{v}}{\partial \Theta}$ can be directly computed using the chain rule. 
\section{Implementation and Parameter Exploration}
\label{sec:implementation}

Our reference implementation is based on \nvdr~\cite{Laine:2020:nvdiffrast} for creating and smoothing the shadow map. The Jacobians $\nabla \tilde{\mv{f}}$ and $\nabla \tilde{\mv{f}^2}$ (w.r.t.\ the scene parameters) are not explicitly computed. Rather, we use reverse-mode differentiation to compute the gradients of an objective function. The relevant derivatives are computed automatically~\cite{Paszke:2019:PyTorch} and are implicitly evaluated.

\begin{figure}
    \centering
    \includegraphics[width=\linewidth]{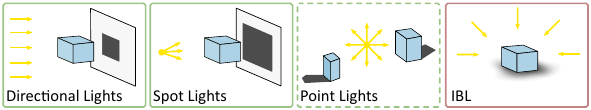}
    \caption{We implement directional lights and spot lights in our framework. An extension to omnidirectional point lights is straight forward. Image-based lighting (IBL) is not supported, but the illumination from some environment maps can be sufficiently well approximated by a collection of point lights.}
    \label{fig:light_support}
\end{figure}

We currently use a single shadow map per light source, which supports directional lights and spot lights. Omnidirectional point lights can be implemented by creating and querying several shadow maps. Area lights or environment lighting would have to be approximated by point light sources -- we leave a more accurate treatment for future work. Figure~\ref{fig:light_support} illustrates the supported light sources in our current implementation.

By building on an existing differentiable rasterizer and automatic differentiation, our code becomes rather straightforward. 
The pseudo-code below is provided for a single light source -- keep in mind that the methods have to be executed for each light source. The pre-computation of the pre-filtered depth images $\mv{m}_1$ and $\mv{m}_2$ is described in Alg.~\ref{alg:shadow_pass} below. Here, we assume that the \textsc{RenderScene} function returns the smoothed discrete depth image $\tilde{\mv{f}}$ as well as $\tilde{\mv{f}^2}$. 
The modular design of \nvdr{} implements this in two steps, first providing the discrete values $\mv{f}$ and $\mv{f}^2 = \mv{f} \otimes \mv{f}$, which are then smoothed using the \texttt{antialias} function. We implement the filtering as a convolution with a kernel of size $k$ (e.g. a Gaussian).
Once the discrete smoothed shadow maps have been computed, the visibility for a point $\mv{x} \in \mathcal{X}$ can be computed with Alg.~\ref{alg:cap}.

{
\setlength{\intextsep}{10pt}%
\begin{algorithm}
\caption{Render the shadow map for a light source $l$}\label{alg:shadow_pass}
\small
\begin{algorithmic}
\Function{RenderShadowMap}{$\mathcal{X}$, $\mv{p}_l$, $k$}
    \State $\tilde{\mv{f}}, \tilde{\mv{f}^2} \gets \textsc{RenderScene}(\mathcal{X}, \mv{p}_l)$
    \State $\mv{m}_1 \gets \textsc{Filter}(\tilde{\mv{f}}, k)$
    \State $\mv{m}_2 \gets \textsc{Filter}(\tilde{\mv{f}^2}, k)$
    \State \Return $\mv{m}_1$, $\mv{m}_2$
\EndFunction
\end{algorithmic}
\end{algorithm}
\begin{algorithm}
\caption{Compute the visibility between a point $\mathbf{x}$ and a light at $\mv{p}$ given the pre-filtered shadow map}\label{alg:cap}
\small
\begin{algorithmic}
\Function{ComputeVisibility}{$\mathbf{x}$, $\mv{p}$, $\mv{m}_1$, $\mv{m}_2$}
\State $\mv{u} \gets L(\mv{x})$
\State $m_1, m_2 \gets \textsc{SampleShadowMap}(\mv{u}, \mv{m}_1,\mv{m}_2)$
\State $d \gets d(\mv{x},\mv{p}), \quad \mu \gets m_1, \quad \sigma^2 \gets m_2 - m_1^2$ %
\If {$d \le \mu$}\ \Return $1$
\Else \ \Return $\sigma^2 / (\sigma^2 + (d-\mu)^2)$
\EndIf
\EndFunction
\end{algorithmic}
\end{algorithm}
}

\begin{figure}
    \centering
    \includegraphics[width=\linewidth]{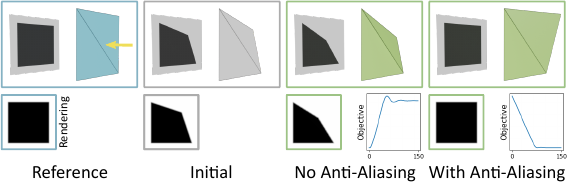}
     \caption{Convergence behavior for a simple scene with and without using a smoothed shadow map in the differentiable rasterizer. Without smoothing, shadow discontinuities are not properly differentiated and optimization fails. 
     }
    \label{fig:minimal_plane}
\end{figure}

\paragraph{Optimization without Smoothed Shadow Map.}

We test our hypothesis that the discrete shadow map $\mv{f}$ and $\mv{f}^2$ must be turned into smooth versions $\tilde{\mv{f}}$ and $\tilde{\mv{f}^2}$ for properly tracking shadow discontinuities. This is facilitated by the modularity of \nvdr{} that allows disabling the smoothing. 

Our setup is as simple as possible: a single square occluder casts a single shadow on a single square receiver. We optimize the position of one displaced vertex of the occluder to reproduce the shadow in a reference configuration (see Figure~\ref{fig:minimal_plane}).
The optimization minimizes the mean squared error to the reference shadow image.

We find that the target configuration is not attained when using the discontinuous discrete shadow map for generating the visibility (i.e., without using \nvdr's \texttt{antialias} function). Importantly, this result is independent of the kernel $k$ used for pre-filtering to generate the moments $\mv{m}_1$ and $\mv{m}_2$. In other words, filtering $\mv{f}$ and $\mv{f}^2$ cannot recover the connection between the scene parameters and the shadow boundaries. Using the smooth functions (i.e., applying \texttt{antialias} to $\mv{f}$ and $\mv{f}^2$) shows the expected  convergence across variations of all other parameters.

\paragraph{Shadow Map Resolution and Filter Kernel.}

\begin{figure}
    \centering
    \includegraphics[width=\linewidth]{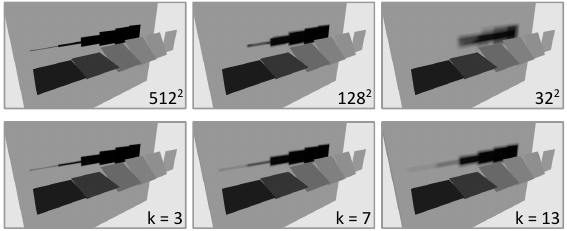}
     \caption{Effect of shadow map resolution (top row) and filter size $k$ (bottom row) on the shadows cast by differently slanted planes. 
     }
    \label{fig:shadow_map_res_and_smoothing}
\end{figure}

\begin{figure}
    \centering
    \includegraphics[width=\linewidth]{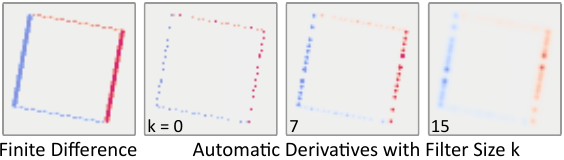}
    \caption{Jacobian of the visibility $\nabla_x \mv{v}$, where $x$ is the horizontal translation of a highly tessellated occluder over a planar receiver.}
    \label{fig:minimal_plane_jacobian}
\end{figure}

The shadow map resolution and the filter kernel size are key parameters as they directly affect the quality of the shadow approximation (see Figure~\ref{fig:shadow_map_res_and_smoothing}).
Instead of investigating the \emph{forward} pass of shadow map rendering, we focus on the \emph{backward} pass, because it influences the gradient computation and any optimization based on it. 

\begin{figure}
    \centering
    \includegraphics[width=\linewidth]{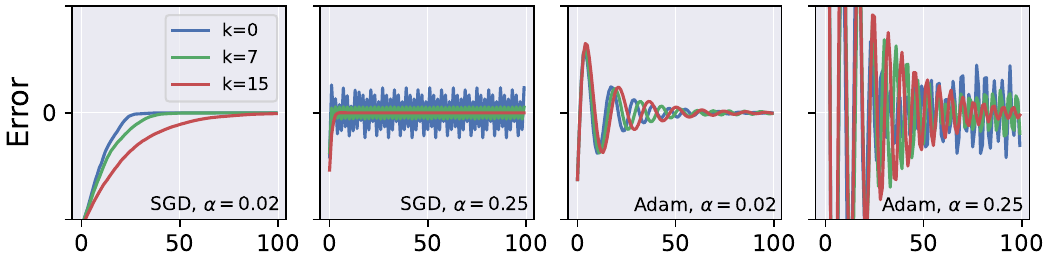}
    \caption{Convergence plots for the single rectangular occluder experiment with different filter kernel sizes $k$. Stochastic gradient descent (left) and \textsc{Adam}~\cite{Kingma:2015:Adam} (right) with small and large steps sizes $\alpha$. The $y$-axis shows the signed error of the $x$ translation.}
    \label{fig:minimal_plane_convergence}
\end{figure}

Consider a scene similar to the one above, now with a highly tessellated and slightly rotated occluder. High tessellation (relative to shadow map pixels) results in sparse shadow map Jacobians for occluder geometry because the anti-aliasing misses primitive boundaries. This might occur for arbitrary differentiable renderers relying on explicit boundary detection.
The filter applied to recover the moments (Eqs.~\eqref{eq:moment1}, \eqref{eq:moment2}) is also applied to the Jacobians of the shadow map and thus affects the Jacobian of the visibility function (see Figure~\ref{fig:minimal_plane_jacobian}). While the filter kernel size is merely a visual effect in the forward pass, it can indeed affect the optimization: Figure~\ref{fig:minimal_plane_convergence} shows that larger filter kernels increase the robustness of gradient descent, both for stochastic gradient descent and momentum-based variants. However, larger kernels also increase the runtime.

Low shadow map resolution may lead to occluders being missed completely by the visibility test
(see Figure~\ref{fig:shadow_map_res_and_smoothing}, top row). These issues can be alleviated by increasing the shadow map resolution; again at the cost of runtime.

\begin{figure}
    \centering
    \includegraphics[width=\linewidth]{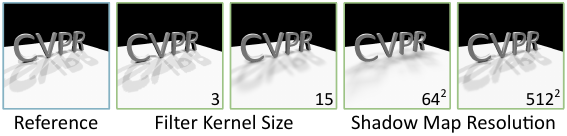}
     \caption{Light direction estimation with four lights, varying filter kernel sizes and shadow map resolutions.}
    \label{fig:light_pose}
\end{figure}

\begin{figure}
    \centering
    \includegraphics[width=\linewidth]{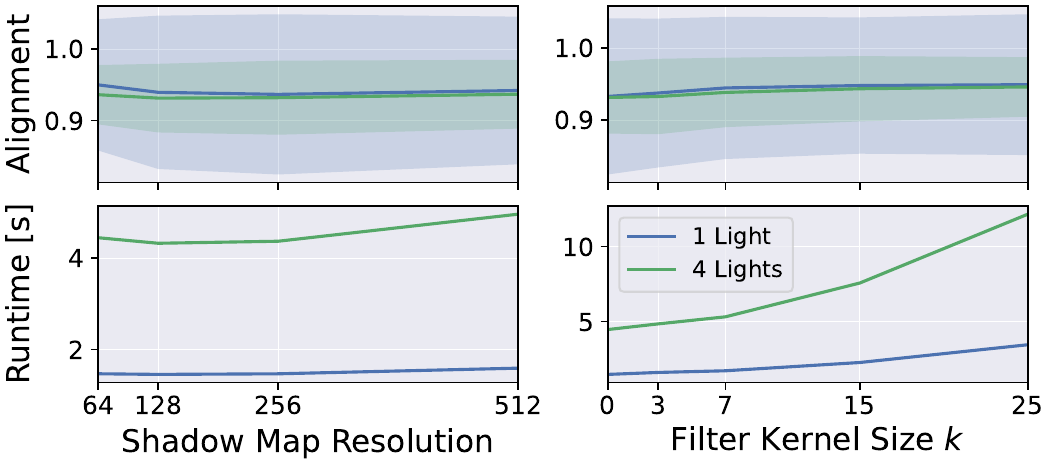}
     \caption{Quantitative results for the parameter exploration, using a light direction estimation experiment. The $y$-axis in the top row is the average alignment of predicted light directions to the reference light directions, measured by the dot product. 
    }
    \label{fig:light_pose_shadow_res_and_kernel_size}
\end{figure}

In the following we consider a more complex setting to test different shadow map resolutions and filter kernel sizes in a more involved optimization: an object is placed on a rectangular floor plane and illuminated by $n$ directional lights (Figure~\ref{fig:light_pose}). Given a single reference image $I_\text{ref}$, 
we wish to recover the $n$ light directions $\{\mathbf{l}_{i}\}$, minimizing the mean squared image error. 

We run the optimization for different objects and varying number of lights $n$. We report qualitative results in Figure~\ref{fig:light_pose} and quantitative results in Figure~\ref{fig:light_pose_shadow_res_and_kernel_size}. As expected, increasing the shadow map resolution or the filter kernel size increases the runtime. The optimization behaves robustly with high accuracy and consistent error bounds across resolutions and filter sizes. Again, the accuracy improves slightly with larger filter size, albeit this improvement seems marginal considering the error bounds. Lowering the shadow map resolution has a similar effect because the effective filter footprint increases if $k$ stays constant. We observe that the optimization for one light can get stuck in local minima where the shadows do not align but the shade of the floor plane matches that in the reference image. For multiple lights, shadows dominate the image, so this scenario is less likely; this could explain the different error bounds. Conversely, recovering multiple lights is more difficult, hence the lower average accuracy. These challenges are specific to the task and, in general, we observe that our method works robustly with different sets of parameters.

\section{Applications}
\label{sec:applications}

We run all experiments on a workstation with an NVIDIA 1070 Ti GPU with 8 GB VRAM. 

\subsection{Monocular Pose Estimation}
\label{sec:application_pose_estimation}

\begin{figure}
    \centering
    \includegraphics[width=\linewidth]{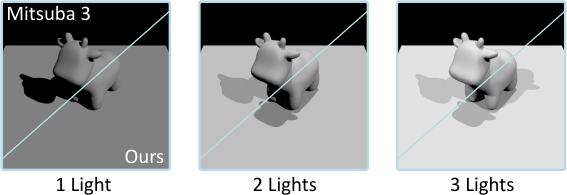}
    \caption{Visual difference between path-traced images by Mitsuba 3~\cite{Jakob:2022:Mitsuba3} in direct illumination mode and our approximation based on local shading models and pre-filtered shadow maps.}
    \label{fig:mitsuba_reproduced}
\end{figure}

Estimating the orientation and position of an object from images is a classic computer vision task. The problem becomes easier for sequences of images~\cite{Fua:3dtracking} or if priors apply, such as in pose estimation of human body parts~\cite{SARAFIANOS20161,deLaGorce:2008:DEODR}. Given only a single (monocular) image makes retrieving the rigid transformation of an known object ill-posed. Shafer and Kanade~\cite{SHAFER1983145} noticed that the shadow of an object in the image provides viable cues for reducing the ambiguity. We consider a simplified version of the 6D problem, allowing only 2 translational and one rotational degree of freedom. We place an object in front of a receiver plane and a directional light orthogonal to the plane. In this setting, we recover the parameters $\mathbf{t} = (x, y)$ (translation parallel to the plane) and $\varphi$ (rotation around the object-centric ``up" axis), given a reference image $I_\text{ref}$.

\begin{figure}
    \centering
    \includegraphics[width=\linewidth]{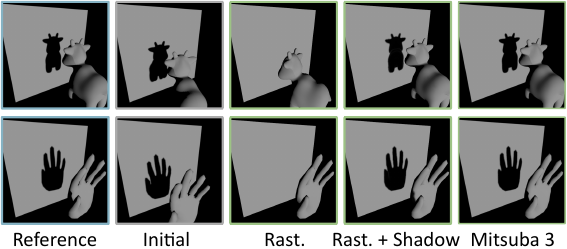}
     \caption{Pose estimation results with path-traced references. In most cases, shadows provide viable cues for convergence (top row) but occasionally pure local shading is sufficient (bottom row).}
    \label{fig:pose_estimation}
\end{figure}

We perform tests across a variety of common 3D mesh models.
The reference images are generated
using the differentiable path tracer Mitsuba 3~\cite{Jakob:2022:Mitsuba3} in direct illumination mode. We verify that our local shading model can generate images similar to the reference images (Figure~\ref{fig:mitsuba_reproduced}).
The results are generated by averaging over several random starting conditions for each object. This experiment is repeated for different camera image resolutions ($128^2, 512^2, 1024^2$).

\begin{table}[t]
	\centering
	\caption{Quantitative results for the pose estimation experiments for different scenes and image resolution $512 \times 512$ (full table in the supplementary material). We measure the rotation error $\Delta \varphi$, the translation error $\Delta \mathbf{t}$, and the total runtime $t$, averaged over ten runs. We include results for a GPU with ray-tracing cores (``RT") and for a setting where we use our renderer as reference (``Our Ref."). Best scores in {\bftab bold}, second best scores \underline{underlined}.}
	\label{tab:pose_estimation_quantitative}
    \setlength{\tabcolsep}{4pt}
	\resizebox{\linewidth}{!}{%
	\begin{tabular}{@{}lrrrrrrrrr@{}} 
		\toprule
        &
        \multicolumn{3}{c}{Mitsuba 3~\cite{Jakob:2022:Mitsuba3}} & \multicolumn{3}{c}{Rast. + Shadows (Ours)} & \multicolumn{3}{c}{Rasterizer} \\
        \cmidrule(lr){2-4} \cmidrule(lr){5-7} \cmidrule(l){8-10}
        & $\downarrow$ $\Delta \varphi [^\circ]$  & $\downarrow \Delta \mathbf{t}$ & $\downarrow$ t [s] & $\downarrow$ $\Delta \varphi [^\circ]$ & $\downarrow \Delta \mathbf{t}$ & $\downarrow$ t [s] & $\downarrow$ $\Delta \varphi [^\circ]$ & $\downarrow \Delta \mathbf{t}$ & $\downarrow$ t [s] \\
        \midrule

        Bunny &  {\bftab 0.23} & {\bftab 0.10} & 325.29 & 0.33 & \underline{0.22} & \underline{3.76} & \underline{0.31} & 0.26 & {\bftab 1.58}  \\
        Dragon & {\bftab 2.86} & {\bftab 3.91} & 306.46 & \underline{3.32} & \underline{4.53} & \underline{3.64} & 8.02 & 11.63 & {\bftab 1.54} \\
        Hand & {\bftab 1.41} & {\bftab 5.18} & 317.40 & \underline{1.68} & \underline{5.36} & \underline{3.66} & 1.85 & 5.66 & {\bftab 1.51} \\
        Spot & {\bftab 0.03} & {\bftab 0.02} & 381.80 & \underline{0.05} & \underline{0.05} & \underline{3.76} & 5.80 & 23.15 & {\bftab 1.77} \\
        \vspace{-0.2cm} & & & & & & & & & \\
        Spot (RT) & {\bftab 0.03} & {\bftab 0.02} & 53.26 & \underline{0.05} & \underline{0.05} & \underline{1.93} & 6.92 & 21.52 & {\bftab 1.05} \\
        Spot (Our Ref.) & \underline{0.05} & \underline{0.12} & 384.24 & {\bftab 0.02} & {\bftab 0.06} & \underline{3.60}  & 5.83 & 23.12 & {\bftab 1.71} \\
        
		\bottomrule
	\end{tabular}
	}
\end{table}

We report quantitative results for the $512^2$ resolution in Table~\ref{tab:pose_estimation_quantitative} (more data is in the supplementary material) and qualitative results in Figure~\ref{fig:pose_estimation}. Mitsuba 3 recovers the rotation and translation most accurately, which \emph{could} be attributed to its accurate gradient computation, using multiple sub-pixel samples.
It should be expected, however, that the results are biased towards the method used for generating the reference images. We show that this is the case by using our renderer as reference for one experiment (see Table~\ref{tab:pose_estimation_quantitative}) and finding our method to yield the most accurate results in this instance. Across all experiments, not using shadows at all and relying purely on the shading of the object surface performs worst, yet still manages to converge to reasonable solutions in some instances.

Unsurprisingly, rasterization is magnitudes faster than path tracing, both with and without shadows. Shadow mapping roughly doubles the runtime in rasterization, however, the pose estimation accuracy is significantly increased: our method consistently yields accuracy similar or close to the path tracing approach. 
Modern GPUs posses hardware units for ray-triangle intersections that significantly increase ray tracing performance. We include scores for a system with such a GPU (NVIDIA RTX 3080 Ti). It improves the path tracing runtime but, at least in this example, rasterization is still $27\times$ faster.

\subsection{Face Reconstruction from Profile Shadows}
\label{sec:application_face_reconstruction}

\begin{figure}
    \centering
    \includegraphics[width=\linewidth]{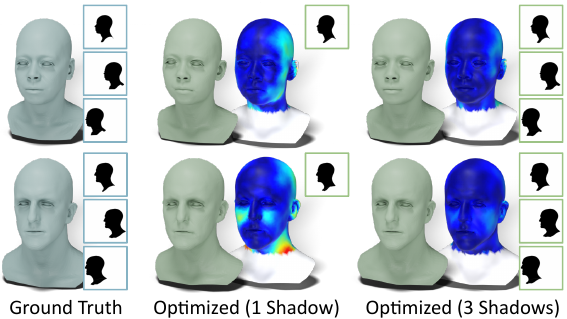} 
    \caption{Face reconstruction from synthetic shadows.}
    \label{fig:face_reconstruction}
\end{figure}

While the variation within certain classes of objects is governed by pose, such as in articulated objects, deformations are best described by morphable models. A prominent example for morphable models is face geometry~\cite{Egger:morphable-face-models}. Our idea is to reconstruct face geometry from one (or more) shadows of the face. We base the experiment on the publicly available morphable model assembled by Li et al.~\cite{Li:2020:FaceMM} (other models would work as well). 

In the first experiment we generated synthetic shadows from the morphable model. This allows comparing the optimization result not just based on images but also geometry (see Figure ~\ref{fig:face_reconstruction}). We find that using more shadows only slightly increases the runtime (the time is not linear in the number of light sources, because we have one base pass and multiple shadow passes) but significantly improves the accuracy (see supplementary material for quantitative results).

\begin{figure}
    \centering
    \includegraphics[width=\linewidth]{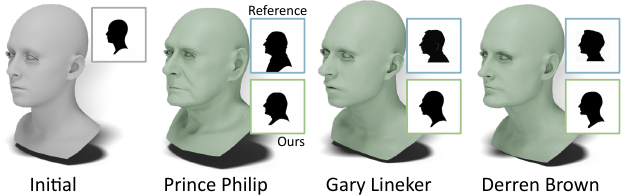} 
    \caption{Face reconstruction from real silhouettes of public figures. We masked out hair, facial hair, and the base.}
    \label{fig:face_reconstruction_celebrity}
\end{figure}

We have also performed reconstructions of face geometry from `real' silhouettes, created manually by artist Charles Burns using scissors and black paper~\cite{burns2011mastering} (the process is shown here~\cite{burns:youtube}). We ignore regions appearing to be (facial) hair or torso as those are not captured by the morphable model we use. While the lack of ground truth makes it difficult to quantify the resulting geometry, visually they appear to be reasonable approximations (Figure~\ref{fig:face_reconstruction_celebrity}).

\subsection{Shadow Art}
\label{sec:application_shadow_art}

\begin{figure*}
    \centering
    \includegraphics[width=\linewidth]{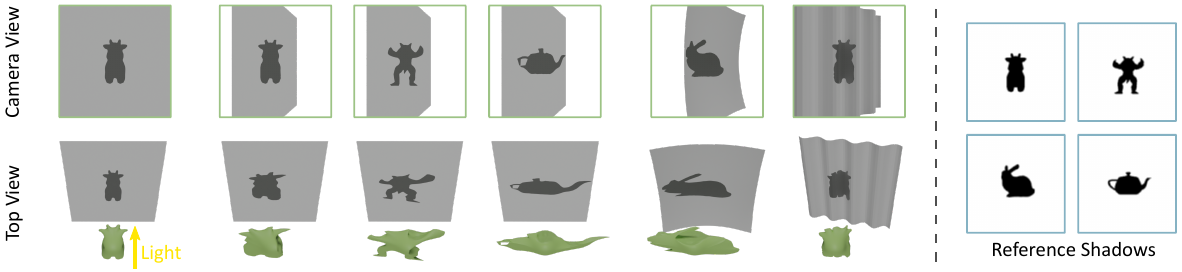} 
    \caption{Shadow Art with one view. Our method can be used for simple scenes with co-located light and camera (left), for more complex settings with perspective cameras that observe the shadow receiver from any direction (middle), and for complex receiver geometry (right).}
    \label{fig:shadow_art_spot_slanted_views}
\end{figure*}

Shadow Art~\cite{Mitra:2009:ShadowArt, Hsiao:2018:WireArt} is the design of a 3D object based only on the shadows it casts. When solving this design problem computationally, it is commonly simplified by assuming that the shadows are cast to planar surfaces orthogonal to directional lights, reducing the shadow images to silhouettes. In the traditional approach~\cite{Mitra:2009:ShadowArt}, the object is represented on a volumetric grid and voxel lines outside the shadow regions are removed -- similar in spirit to visual hulls~\cite{Laurentini:1994:VisualHull} or space carving~\cite{Kutulakos2000}. Sadekar \etal{}~\cite{Sadekar:2022:ShadowArtRevisited} perform Shadow Art with differentiable rendering of meshes,
but the limitation of co-located cameras and lights remains.

In our framework, lights and cameras can be placed arbitrarily. Also, curved shadow receivers work out-of-the-box (Figure~\ref{fig:shadow_art_spot_slanted_views}). The only (minor) constraints we impose are gradient pre-conditioning~\cite{Nicolet:2021:LargeSteps} and a smoothness term between the normals of neighboring faces~\cite{Luan:2021:Unified, Worchel:2022:NDS} to avoid arbitrary vertex movements during optimization.

\begin{figure}
    \centering
    \includegraphics[width=\linewidth]{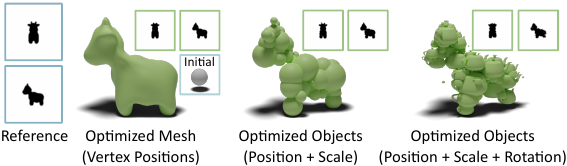} 
    \caption{Shadow art with two views. Our method can be used in any setting that permits differentiable rendering, for example optimizing vertex positions of a mesh (left) or optimizing the parameters of multiple geometric objects (middle and right). 
    }
    \label{fig:shadow_art_2views}
\end{figure}

A single triangle mesh is only one possible representation of the generated geometry. The combination of automatic differentiation and gradient descent allows optimizing arbitrary scene parameters. For example, we can also represent the object as a collection of shapes (see Figure~\ref{fig:shadow_art_2views}). Beyond our reference implementation based on triangle meshes, the same approach readily works with any geometry representation as long as it can be differentiably rendered (\eg{} signed distance functions~\cite{Vicini:2022:DiffSdf, Bangaru:2022:NeuralSDFReparam} or spheres~\cite{Lassner:2021:Pulsar}).

\subsection{Interactive Modeling from Shadows}
\label{sec:application_shadow_modeling}

\begin{figure}
    \centering
    \includegraphics[width=\linewidth]{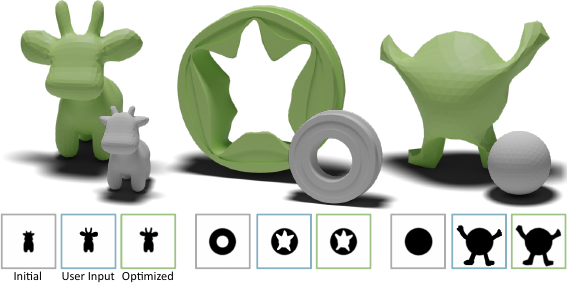}
    \caption{Shadow art from one view, with co-located camera and light is used as an interactive modeling tool. Users can modify the shadow of the reference object (gray) and an appropriate deformation is applied to the mesh in real-time (green).}
    \label{fig:shadow_art_modeling}
\end{figure}

Our rasterization-based implementation generates shadows and -- most importantly -- the gradients of the objective function in real time. We demonstrate this by turning the shadow art setting, which is usually concerned with static reference shadows, into a ``shadow modeling" setting, where a user can interactively paint the target shadows and an appropriate deformation is applied to the initial geometry in real time (see Figure~\ref{fig:shadow_art_modeling}). Note that using co-located camera and light is a design decision since painting silhouettes feels intuitive -- the method is not limited to this setup.

\section{Conclusion}
\label{sec:conclusions}

Our approach to inverse graphics using differentiable rasterization with local shading models and shadow maps helps in closing the gap between efficient differentiable renderers based on rasterization and renderers based on path tracing and Monte Carlo methods that more accurately model light transport.

\paragraph{Limitations}
Focusing on direct shadows neglects other effects of global illumination. But even within the framework of local shading models with shadow maps we have made simplifiying assumptions: so far we consider only point light sources (instead of area lights) and for those we consider only one planar shadow map. Also, the particular choice of variance shadow maps suffers from artifacts, such as \emph{light bleeding}, i.e.\ shadows becoming lighter in regions of high depth variance. 
Independent of the illumination effects, we have so far only considered triangle meshes for the representation of geometry.

\paragraph{Possible Generalizations}

Apart from shadows, various effects of global illumination have been recreated by efficient real-time approximations. Approaches very similar to ours may readily work with ambient occlusion~\cite{10.1145/192161.192244}, reflections and refractions~\cite{10.1145/311535.311554}, or generally indirect illumination. 

Other geometry representations that are popular in the vision community, such as signed distance fields~\cite{Jiang_2020_CVPR} or NeRFs~\cite{mildenhall2020nerf}, should work with shadow maps as well -- although it is not clear if the same efficiency can be achieved, as the fast computation is mainly achieved by rasterizing triangle primitives.

{\small
\bibliographystyle{ieee_fullname}
\bibliography{egbib}
}

\iftoggle{preprint}{%
    \clearpage
    {
        \renewcommand{\appendixpagename}{Supplementary Material} %
        \begin{appendices}
        \section{Implementation and Renderer Details}

We implemented our method in Python\iftoggle{preprint}{\footnote{An open-source implementation is available at \url{https://github.com/mworchel/differentiable-shadow-mapping}}}{}, on top of the automatic differentiation framework PyTorch~\cite{Paszke:2019:PyTorch} and use the differentiable rasterization primitives by Laine \etal{}~\cite{Laine:2020:nvdiffrast} as foundation for our renderer.
Our rendering pipeline follows a deferred shading architecture~\cite{Deering:1988:DeferredShading}, so we first rasterize the scene geometry and then perform a shading pass that computes the light-material interaction. This architecture allows us to consider an arbitrary number of lights. Currently, all surfaces use a BRDF (bidirectional reflectance distribution function) with only the Lambert diffuse term but an extension to more physically-based material models, like the Disney ``principled" BRDF~\cite{Burley:2012:DisneyBRDF} or variants of it~\cite{Karis:2013:RealShadingUE, Lagarde:2014:Frostbite}, is trivial.

For rendering the shadow maps, we use the same rasterization primitives as for the primary camera. The camera built for each light depends on the light type: (1) a directional light uses a camera with orthographic project that is placed far outside of the scene, facing the origin from the specified direction; (2) a spot light uses a camera that is placed at its position, facing the desired direction, and using a perspective projection with a configurable field of view.

\section{Experiment Details}

Our scenes are roughly centered at the origin and after loading 3D models, we scale them to the $[-1, 1]^3$ cube. This allows us to use consistent near and far planes across experiments, both for the frusta of the actual cameras and the frusta of the light cameras. 

For all experiments -- except those explicitly stating otherwise -- we use a shadow map resolution of $256 \times 256$. We implement two filters for pre-filtering the shadow map: a box filter and a Gaussian filter. We use both filters in the experiments, usually with kernel size $k=3$ or $k=5$. The Gaussian filter produces more pleasing visual results whereas using the box filter is a little faster. Generally, we observed robustness to the exact choice of filter. 

\paragraph{Convergence and Jacobian Experiments.}

We use a ``minimal plane" setup for the convergence and Jacobian experiments, consisting of a single planar shadow receiver and a single planar occluder (both parallel), illuminated by a directional light orthogonal to them. The camera that produces the shadow images used for optimization is located between the receiver and the occluder, focusing the receiver. The highly tessellated and slightly rotated occluder that we use for computing the Jacobian has 130,000 faces. For the simple experiment that investigates the effect of shadow map filtering on the convergence, we displace the occluder horizontally and the goal of the optimization is to recover the original position, only by using the shadow images.

\paragraph{Light Direction Estimation.}

The experiment of light direction estimation has the following general setup: an object is placed on a rectangular floor plane and illuminated by $n$ directional lights. We render a reference image with a fixed camera that observes the object. The goal of the optimization is to recover the target light directions from some initial state, only by comparing the rendered image to the target image. 

We run the experiment for 6 objects in total. For each parameter setting (e.g. filter size) we perform 150 runs of the experiment, with randomized target and initial conditions. We ensure that these configurations are deterministic, so -- as an example -- the optimizations for filter size $k = 3$ and $k = 15$ use the same set of 150 configurations. The average accuracy is computed as an average over the 6 objects and 150 runs; our error bounds are based on the standard deviation, respectively.

The accuracy (alignment) is computed as follows: Given the $n$ target light directions $\{ \mv{l}_i \}$ and the estimated light directions $\{ \mv{l}^\text{opt}_j \}$, we first perform a greedy matching between both sets, trying to maximize the dot product $\mv{l}_i \cdot \mv{l}^\text{opt}_{k_i}$ between a direction $i$ and its match $k_i$. Note that this mapping is bijective, so there is a one-to-one correspondence between directions. We then compute the alignment as the average dot product over all directions:
\begin{equation}
    \frac{1}{n} \sum_i \mathbf{l}_{i} \cdot \mathbf{l}^\text{opt}_{k_i}
\end{equation}

\paragraph{Monocular Pose Estimation.} 

The general setup for the pose estimation application 
is as follows: the scene consists of a single planar receiver that is illuminated by a directional light orthogonal to it. The object whose pose we wish to recover is placed in front of the receiver, such that it casts a shadow. For each optimization run, we first randomly offset the object in the plane parallel to the receiver and randomly rotate it around its up-axis (the random offsets and rotations are deterministic). Then, we recover the original pose in an optimization, which compares rendered images to a reference image that shows the object without offset or rotation. The camera is located at a position where it sees both the object and the shadow receiver plane.

We use Mitsuba 3~\cite{Jakob:2022:Mitsuba3} for creating the reference images and as a baseline in the optimization. Since Mitsuba 3 does not support differentiation for directional lighting, we use a far-away, rectangular area light as a proxy in the optimization. We calibrate this proxy to match the directional light. In our experiments, timings for Mitsuba 3 showed no noticeable dependence on the choice of emitter. 

We use \textsc{Adam}~\cite{Kingma:2015:Adam} for the optimization, with a step size of $0.01$ and $\beta_1 = 0.9$, $\beta_2 = 0.999$ and run $150$ iterations.

\paragraph{Face Reconstruction from Profile Shadows.}

\begin{figure}
    \centering
    \includegraphics[width=0.48\linewidth]{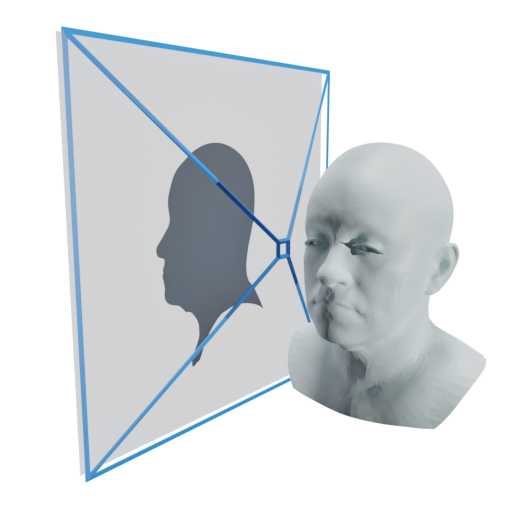}
    \includegraphics[width=0.25\linewidth]{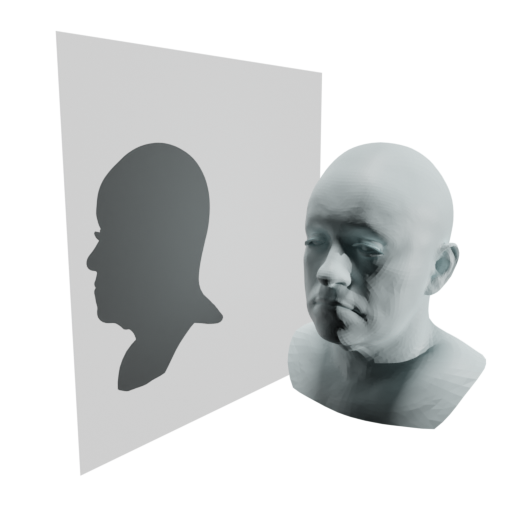}
    \includegraphics[width=0.25\linewidth]{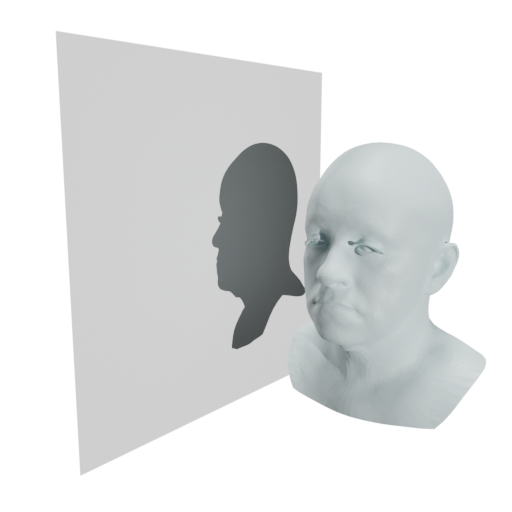}
    \caption{General setup of the face reconstruction experiment. The face model is placed in front of a planar shadow receiver and the camera (blue) is placed between them, focusing the receiver (left). For the experiments with three shadows, we add two additional lights with small horizontal offsets (middle, right).}
    \label{fig:face_reconstruction_setup}
\end{figure}

The basic setup for face reconstruction is similar to that of pose estimation: a single planar shadow receiver is illuminated by a directional light orthogonal to it. The head model is placed in front of the plane so that is casts a shadow. However, here the camera is located between the head and the receiver, focusing the receiver (see Figure~\ref{fig:face_reconstruction_setup}). For the setup with three shadows, we add two additional directional lights.

We use artificial geometry to evaluate the quality of our fit quantitatively. When computing the geometric distance between the reference geometry and our prediction, we only compare the meshes from the neck upwards, because the shadow of the upper torso is not included in the optimization.

In the experiment that uses silhouette images of real public figures, we first scale the images and align them roughly with the shadow observed by the camera in our scene. We also optimize the translation of the head model to reduce the effect of small inaccuracies in this alignment.

We use \textsc{Adam}~\cite{Kingma:2015:Adam} with a step size of $0.01$ for the translation and a step size of $0.2$ for the identity weights used by the morphable face model, both with $\beta_1 = 0.9$, $\beta_2 = 0.999$. We run the optimization for $400$ iterations.

\paragraph{Shadow Art.}

The basic shadow art setup is again similar to the setup for face reconstruction and pose estimation: a single shadow receiver illuminated by an orthogonal directional light with the geometry that is optimized in front of the plane. The experiments with two views use a second shadow receiver that is placed orthogonal to the first one, illuminated by a second directional light orthogonal to it (see Figure~\ref{fig:shadow_art_setup}). For each receiver plane, we place an orthographic camera between the object and the plane: these are the cameras producing the shadow images used in the optimization. 

In selected experiments, we use a perspective camera that observes the receiver plane from the side, demonstrating that light rays and view rays do not need to coincide with our framework. We also replace the planar shadow receiver with a curved receiver to show that the receiver does not necessarily need to be flat. 
In our demo code, we include a shadow art sample, which reproduces these settings for a single view.

The shadow art experiments that deform a mesh start with a sphere of 12,800 triangles and finish in roughly 15 seconds for each object. We run 400 iterations with \textsc{Adam}~\cite{Kingma:2015:Adam}, using a step size of $0.2$, $\beta_1 = 0.9$, $\beta_2 = 0.999$, $\lambda=20$ for gradient preconditioning~\cite{Nicolet:2021:LargeSteps}, and a weight of $0.2$ for the normal consistency regularization.

The shadow art experiments that optimize the translations, scales, and rotations of an object collection initially start with one instance of the object and duplicate the number of objects every 150 iterations (each object spawns two objects with small offsets to avoid interpenetration). We run 1100 iterations with \textsc{Adam}~\cite{Kingma:2015:Adam}, using a step size of $0.005$, $\beta_1 = 0.9$, $\beta_2 = 0.999$.

\begin{figure}
    \centering
    \includegraphics[width=0.49\linewidth, trim=0 50 50 50]{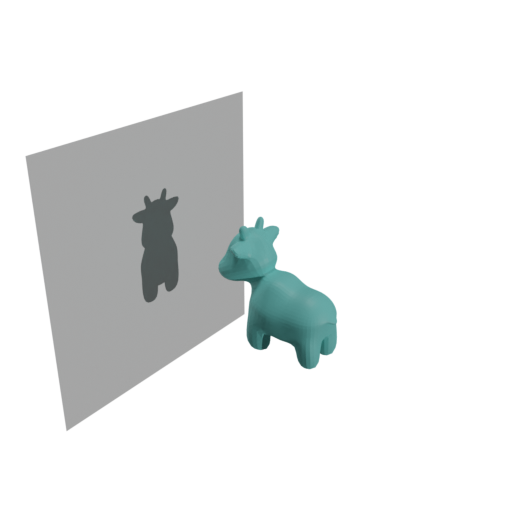}
    \includegraphics[width=0.49\linewidth, trim=50 50 0 50]{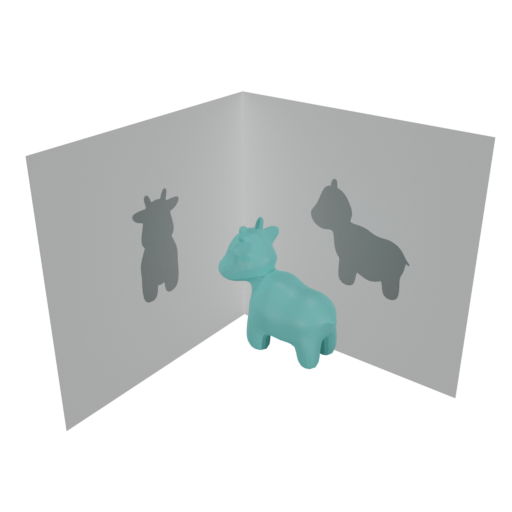}
    \caption{Shadow art setup for one view (left) and two views (right), with a reference object.}
    \label{fig:shadow_art_setup}
\end{figure}

\paragraph{Interactive Modeling from Shadows.}

The interactive modeling setup is equivalent to the one-view shadow art setup. In contrast to shadow art, the intend here is to modify an \emph{existing} model based on user interaction with the shadow image. We first load an input mesh, generate a shadow image from the unmodified mesh, and write the image to the disk. The optimization loop then reads back this shadow image in each iteration and uses it as a target image to drive the model deformation.

\section{Additional Results}

\paragraph{Monocular Pose Estimation.}

Table~\ref{tab:pose_estimation_quantitative} contains the quantitative results for all resolutions ($128^2$, $512^2$, $1024^2$) as well as time measurements per iteration. We can perform approximately 70 ($128^2$), 40 ($512^2$), and 20 ($1024^2$) optimization steps per second, which allows following the optimization in real-time.

\begin{table*}[t]
	\centering
	\caption{Quantitative results for the pose estimation experiments for different scenes and resolutions. We measure the rotation error $\Delta \varphi$, the translation error $\Delta \mathbf{t}$, the total runtime, and runtime per iteration, averaged over ten runs. We include results for a GPU with ray-tracing cores (``RT") and for a setting where we use our renderer as reference (``Our Ref."). Best scores in {\bftab bold}, second best scores \underline{underlined}.}
	\label{tab:pose_estimation_quantitative}
    \setlength{\tabcolsep}{4pt}
	\resizebox{0.8\linewidth}{!}{%
	\begin{tabular}{@{}lrrrrrrrrr@{}} 
		\toprule
        &
        \multicolumn{3}{c}{Mitsuba 3~\cite{Jakob:2022:Mitsuba3}} & \multicolumn{3}{c}{Rasterizer + Shadows (Ours)} & \multicolumn{3}{c}{Rasterizer} \\
        \cmidrule(lr){2-4} \cmidrule(lr){5-7} \cmidrule(l){8-10}
        & $\downarrow$ $\Delta \varphi [^\circ]$  & $\downarrow \Delta \mathbf{t}$ & $\downarrow$ time [s] (/it) & $\downarrow$ $\Delta \varphi [^\circ]$ & $\downarrow \Delta \mathbf{t}$ & $\downarrow$ time [s] (/it) & $\downarrow$ $\Delta \varphi [^\circ]$ & $\downarrow \Delta \mathbf{t}$ & $\downarrow$ time [s] (/it) \\
        \midrule
        
        \multicolumn{1}{@{}l@{}}{$128 \times 128$} & & & & & & & & & \\
        \cmidrule(r){1-1}
        Bunny & {\bftab 0.31} & {\bftab 0.05} & 54.13 (0.361) & \underline{0.69} & \underline{0.27} & \underline{2.27} (\underline{0.015}) & 3.04 & 5.56 & {\bftab 1.06} ({\bftab 0.007})\\
        Dragon & {\bftab 2.78} & \underline{4.01} & 52.46 (0.349) & \underline{3.32} & {\bftab 3.94} & \underline{2.09} (\underline{0.014}) & 8.01 & 14.31 & {\bftab 1.03} ({\bftab 0.007}) \\
        Hand & {\bftab 1.44} & \underline{5.24} & 53.17 (0.354) & \underline{3.66} & {\bftab 4.02} & \underline{2.08} (\underline{0.014}) & 6.80 & 12.42 & {\bftab 1.03} ({\bftab 0.007}) \\
        Spot & {\bftab 0.02} & {\bftab 0.11} & 58.19 (0.388) & \underline{7.40} & \underline{12.83} & \underline{2.11} (\underline{0.014}) & 12.47 & 44.70 & {\bftab 1.05} ({\bftab 0.007}) \\\\
        
        \multicolumn{1}{@{}l@{}}{$512 \times 512$} & & & & & & & & & \\
        \cmidrule(r){1-1}
        Bunny &  {\bftab 0.23} & {\bftab 0.10} & 325.29 (2.168) & 0.33 & \underline{0.22} & \underline{3.76} (\underline{0.024}) & \underline{0.31} & 0.26 & {\bftab 1.58} ({\bftab 0.009}) \\
        Dragon & {\bftab 2.86} & {\bftab 3.91} & 306.46 (2.043) & \underline{3.32} & \underline{4.53} & \underline{3.64} (\underline{0.023}) & 8.02 & 11.63 & {\bftab 1.54} ({\bftab 0.009}) \\
        Hand & {\bftab 1.41} & {\bftab 5.18} & 317.40 (2.116) & \underline{1.68} & \underline{5.36} & \underline{3.66} (\underline{0.023}) & 1.85 & 5.66 & {\bftab 1.51} ({\bftab 0.009}) \\
        Spot & {\bftab 0.03} & {\bftab 0.02} & 381.80 (2.545) & \underline{0.05} & \underline{0.05} & \underline{3.76} (\underline{0.023}) & 5.80 & 23.15 & {\bftab 1.77} ({\bftab 0.010})\\
        \vspace{-0.2cm} & & & & & & & & & \\
        Spot (RT) & {\bftab 0.03} & {\bftab 0.02} & 53.26 (0.354) & \underline{0.05} & \underline{0.05} & \underline{1.93} (\underline{0.012}) & 6.92 & 21.52 & {\bftab 1.05} ({\bftab 0.006}) \\
        Spot (Our Ref.) & \underline{0.05} & \underline{0.12} & 384.24 (2.561) & {\bftab 0.02} & {\bftab 0.06} & \underline{3.60} (\underline{0.022}) & 5.83 & 23.12 & {\bftab 1.71} ({\bftab 0.010}) \\\\
        
        \multicolumn{1}{@{}l}{$1024 \times 1024$} & & & & & & & & & \\
        \cmidrule(r){1-1}
        Bunny & {\bftab 0.22} & {\bftab 0.12} & 1245.60 (8.304) & \underline{0.31} & \underline{0.21} & \underline{7.72} (\underline{0.047}) & 0.33 & 0.25 & {\bftab 3.71} ({\bftab 0.021}) \\
        Dragon & {\bftab 2.86} & {\bftab 3.87} & 1151.74 (7.678) & \underline{3.22} & \underline{4.48} & \underline{7.53} (\underline{0.046}) & 7.85 & 11.48 & {\bftab 3.18} ({\bftab 0.018}) \\
        Hand & {\bftab 1.42} & {\bftab 5.38} & 1200.74 (8.005) & \underline{1.61} & \underline{5.43} & \underline{7.72} (\underline{0.048}) & 1.76 & 5.66 & {\bftab 3.13} ({\bftab 0.018}) \\
        Spot & {\bftab 0.03} & {\bftab 0.03} & 1441.56 (9.610) & {\bftab 0.03} & \underline{0.04} & \underline{7.64} (\underline{0.046}) & \underline{2.58} & 3.93 & {\bftab 3.19} ({\bftab 0.017}) \\
		\bottomrule
	\end{tabular}
	}
\end{table*}

\paragraph{Face Reconstruction from Profile Shadows.}

\begin{figure}
    \centering
    \includegraphics[width=\linewidth]{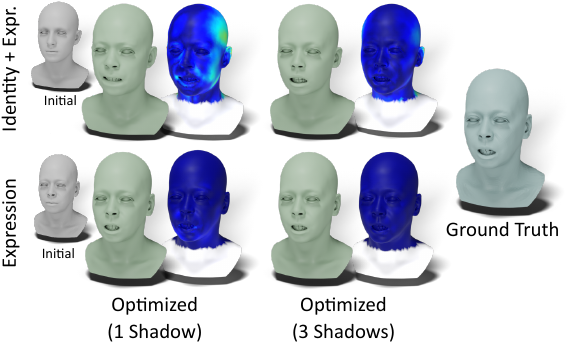}
    \caption{We use a morphable face model that allows controlling the face identity and the expression. We optimize both (top row) and only the expression (bottom row) based on the shadows cast by the model.}
    \label{fig:face_reconstruction_expression}
\end{figure}

Table~\ref{table:face_reconstruction} contains the quantitative results for four artificial faces models. In the main paper, we only optimize the identity weights of the morphable face model, which effectively determine the fundamental \emph{shape} of the face. We also experimented with optimizing the identity and expression as well as only the expression, while keeping the identity fixed (see Figure~\ref{fig:face_reconstruction_expression}). The latter might be useful if the shape of the head is known (\eg{} because it was scanned previously) and the task is to recover facial expressions using shadows, for example in the context of face motion capture.

\begin{table}[t]
	\centering
	\caption{Quantitative results for face reconstruction from shadows, using artificial head geometry generated with a morphable face model. Best scores in \textbf{bold}.}
	\label{table:face_reconstruction}
    \setlength{\tabcolsep}{4pt}
	\resizebox{\columnwidth}{!}{%
	\begin{tabular}{@{}lrrrrrr@{}} 
		\toprule
        &
        \multicolumn{3}{c}{1 Shadow} & \multicolumn{3}{c}{3 Shadows} \\
        \cmidrule(lr){2-4} \cmidrule(l){5-7}
        Face ID & $\downarrow D_\text{mean}$ & $\downarrow D_\text{Hausdorff}$  & $\downarrow$ time [s] (/it) & $\downarrow D_\text{mean}$ & $\downarrow D_\text{Hausdorff}$ & $\downarrow$ time [s] (/it) \\
        \midrule
        1 & 0.011 & 0.114 & {\bftab 18.86} ({\bftab 0.047}) & {\bftab 0.006} & {\bftab 0.066} & 27.11 (0.068) \\
        6 & 0.012 & 0.108 & {\bftab 19.00} ({\bftab 0.048}) & {\bftab 0.010} & {\bftab 0.106} & 27.32 (0.068) \\
        8 & 0.014 & 0.129 & {\bftab 19.36} ({\bftab 0.048}) & {\bftab 0.007} & {\bftab 0.054} & 27.37 (0.068) \\
        131 & 0.014 & 0.069 & {\bftab 18.09} ({\bftab 0.045}) & {\bftab 0.008} & {\bftab 0.046} & 27.46 (0.069) \\
        \midrule
        mean & 0.013 & 0.105 & {\bftab 18.83} ({\bftab 0.047}) & {\bftab 0.008} & {\bftab 0.068} & 27.32 (0.068) \\
		\bottomrule
	\end{tabular}
	}
\end{table}

\section{Limitation: Light Bleeding}

\begin{figure}
    \centering
    \includegraphics[width=\linewidth]{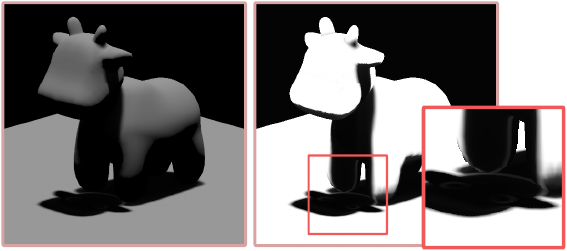}
    \caption{Shaded image (left) and visibility (right) with visible light bleeding artifact.}
    \label{fig:light_bleeding}
\end{figure}

It is well known that Variance Shadow Maps~\cite{Donnelly:2006:VSM} -- our particular choice of pre-filtered shadow maps -- suffer from light bleeding in regions with high depth variance (see Figure~\ref{fig:light_bleeding}). While we noticed no immediate impact in our experiments, it might be desirable to reduce light bleeding by using heuristics~\cite{Nguyen:2007:GpuGems3} or implementing other pre-filtered shadow maps~\cite{Peters:MSM, Annen:CSM, Annen:ESM} in our framework.
        \end{appendices}
    }
}{%
}

\end{document}